\documentclass[useAMS,usenatbib]{mn2e}
\usepackage{graphicx}

\title[Escape of CR electrons from SNRs]{Escape of cosmic-ray electrons from supernova remnants}
\author[Y. Ohira et al.]{Yutaka Ohira$^{1}$\thanks{E-mail:ohira@phys.aoyama.ac.jp}, Ryo Yamazaki$^{1}$, Norita Kawanaka$^{2}$ and Kunihito Ioka$^{3,4}$\\
$^{1}$Department of Physics and Mathematics, Aoyama Gakuin University, 
5-10-1 Fuchinobe, Sagamihara 252-5258, Japan \\
$^{2}$Racah Institute of Physics, Hebrew University of Jerusalem, Jerusalem, 91904, Israel \\
$^{3}$Theory Center, Institute of Particle and Nuclear Studies, KEK, Oho 1-1, Tsukuba 305-0801, Japan \\
$^{4}$The Graduate University for Advanced Studies (Sokendai), Oho 1-1, Tsukuba 305-0801, Japan
}
\begin{document}

\date{Accepted 2012 August 9. Received 2012 August 7; in original form 2012 March 19}

\pagerange{\pageref{firstpage}--\pageref{lastpage}} \pubyear{2012}

\maketitle

\label{firstpage}

\begin{abstract}
We investigate escape of cosmic ray (CR) electrons from a
supernova remnant (SNR) to interstellar space. 
We show that CR electrons escape in order from high energies to low energies like CR 
nuclei, while the escape starts later than the 
beginning of the Sedov phase at an SNR age of $10^3-7\times10^3~{\rm yrs}$ 
and the maximum energy of runaway CR electrons 
is below the knee about $0.3-50~{\rm TeV}$ because unlike CR nuclei, 
CR electrons lose their energy due to synchrotron radiation.
Highest energy CR electrons will be directly probed by 
AMS-02, CALET, CTA and LHAASO experiments, or have been already 
detected by H.E.S.S. and MAGIC as a cutoff in the CR electron spectrum. 
Furthermore, we also calculate the spatial distribution of runaway CR 
electrons and their radiation spectra around SNRs. 
Contrary to common belief, maximum-energy photons of synchrotron 
radiation around $1~{\rm keV}$ are emitted by runaway CR electrons which 
have been caught up by the shock. 
Inverse Compton scattering by runaway CR electrons can dominate 
the gamma-ray emission from runaway CR nuclei via pion decay, 
and both are detectable by CTA and LHAASO as clues to the CR origin
and the amplification of magnetic fluctuations around the SNR.
We also discuss middle-aged and/or old SNRs as unidentified very-high-energy gamma-ray sources.
\end{abstract}


\section{Introduction}

The origin of cosmic rays (CRs) is a longstanding problem in astrophysics. 
Supernova remnants (SNRs) are thought to be the origin of Galactic CR 
nuclei and electrons.
The most popular SNR acceleration mechanism is the diffusive shock 
acceleration \citep{axford77,krymsky77,bell78,blandford78}.
In fact, {\it Fermi} and {\it AGILE} show that middle-aged SNRs interacting 
with molecular clouds emit GeV gamma rays \citep[e.g.][]{abdo09,tavani10} 
and the origin of the GeV gamma rays can be interpreted as the decay of 
neutral pions produced by CR nuclei \citep[e.g.,][]{ohiraetal11}.
In addition, X-ray observations provide an evidence that electrons are 
accelerated to highly relativistic energies in SNR shocks \citep{koyama95}.
SNRs have been also observed by ground-based Cherenkov telescopes \citep[e.g.][]{muraishi00,aharonian06a}.
However, it is still unclear whether the origin of TeV gamma rays is inverse 
Compton scattering  by CR electrons or the decay of neutral 
pions produced by inelastic collisions between CR protons and 
ambient thermal nuclei. 
In addition, there are many unidentified very-high-energy gamma-ray sources 
in our Galaxy, and their emission mechanism 
is also still unclear.

Escape of CR nuclei from SNRs has been investigated by several authors
\citep[e.g.,][]{ptuskin05,ohiraetal10,caprioli10,ohira11,drury11}, 
and emission from runaway CR nuclei has also been investigated 
\citep{aharonian96,gabici09,ohiraetal11,ohiraetal12,ellison11}.
Runaway CR nuclei can emit gamma rays from the exterior of accelerators 
and its radiation spectrum is softer than that from the interior 
because of energy-dependent diffusion of CRs \citep{aharonian96}. 
The escape process of CRs from accelerators is important because it 
changes the CR spectrum.
The runaway CR spectrum depends not only on the CR spectrum 
at accelerators but also on the evolution of the maximum energy 
and the evolution of the number of accelerated CRs \citep{ohiraetal10}.
In addition, considering the escape effect, one can explain the spectral difference 
between CR protons and helium observed by CREAM \citep{ahn10, ohira11}.

However, the same processes for CR electrons suffering cooling effects have never been
investigated so far. 
Cooling processes via synchrotron radiation and inverse Compton scattering 
are important for CR electrons because their cooling times can be smaller than
other characteristic timescales.
Therefore, it is unclear whether CR electrons can escape from 
SNRs or not, and whether the electron cooling affects the spectrum of 
runaway CRs electrons or not, as in the energy spectrum of CR electrons 
inside the SNRs. 
Runaway CR electrons can produce gamma rays outside accelerators. 
In the low density region, these gamma rays probably dominate over those 
from runaway CR nuclei and may be observed as unidentified 
very-high-energy gamma-ray sources.
This is because CR nuclei can not produce sufficient gamma rays there 
electrons can always produce enough gamma rays 
via inverse Compton scattering of CMB photons \citep[see equation (7) of][]{katz08}. 
Furthermore, distant SNRs can not be easily identified by radio observations
because of brighter emission from molecular clouds than synchrotron radiation.

Direct measurements of the CR electron spectrum
may also bring us important information on escape of CR electrons.
So far, {\it Fermi}, H.E.S.S. and MAGIC have revealed the spectrum of CR electrons 
and positrons up to a few ${\rm TeV}$ \citep{aharonian09,ackermann10,tridon11}.
Future experiments, such as AMS-02, CALET, CTA, and LHAASO, will 
measure the spectrum of CR electrons up 
to $1-100~{\rm TeV}$ with good accuracy 
\citep{kounine10,torii08,cta10,cao10}.
\citet{kobayashi04} pointed out that a few nearby sources like
Vela and Cygnus loop  may leave  their own signatures in the TeV energy band, 
and that we will be able to see a spectral shape of CR electrons and positrons 
from a single source.
Moreover, \citet{kawanaka11} have pointed out that escape of CR
electrons can be investigated by future observations of CR electrons, 
in particular, with a low-energy spectral cutoff.

In this paper, we investigate escape of CR electrons from an SNR. 
Our stance in this paper is that 
because the evolution of the magnetic field has not been understood theoretically,
we predict observable quantities by using phenomenological approaches 
and restrict the phenomenological models by comparing the predicted 
values and observations. 
Assuming the evolution of the maximum energy of CR protons (hereafter 
we treat nuclei as protons) during the Sedov phase \citep{gabici09,ohiraetal10}, 
we obtain the evolution of the diffusion coefficient, the magnetic field, 
and the maximum energy limited by synchrotron cooling (section \ref{sec:2}).
Moreover, we calculate spatial distributions of runaway CR electrons 
around the SNR (section \ref{sec:3}), and radiation spectra from runaway 
CR electrons (section \ref{sec:4}).
Finally, we discuss our results (section \ref{sec:5}).
\section[]{Escape of CR electrons}
\label{sec:2}
\subsection{Evolution of SNRs}
\label{sec:2.1}
In this paper, in order to understand essential features of escape of 
CR electrons, we assume simple evolutions of the shock radius, 
$R_{\rm sh}(t)$, and the shock velocity, $u_{\rm sh}(t)$, as in
the following forms:
\begin{eqnarray}
R_{\rm sh}(t)=R_{\rm S}\times \left\{ \begin{array}{ll}
\left(\frac{t}{t_{\rm S}}\right)                     & ~(~t \leq t_{\rm S}~) \\
\left(\frac{t}{t_{\rm S}}\right)^{\frac{2}{5}} & ~(~t_{\rm S}\leq t~) \\
\end{array} \right. ~~,
\label{eq:rsh}
\end{eqnarray}
\begin{eqnarray}
u_{\rm sh}(t)=\frac{R_{\rm S}}{t_{\rm S}}\times \left\{ \begin{array}{ll}
1                                                               & ~(~t \le t_{\rm S}~) \\
\left(\frac{t}{t_{\rm S}}\right)^{-\frac{3}{5}} & ~(~t_{\rm S}\leq t~) \\
\end{array} \right. ~~,
\label{eq:ush}
\end{eqnarray}
where $t$ is the SNR age, and $R_{\rm S}$ and $t_{\rm S}$ are 
the SNR shock radius and the SNR age at the beginning of the 
Sedov phase, respectively (see Table~\ref{table1} for summary).
The Sedov phase starts when the swept-up mass, 
$4\pi n_{\rm ISM}m_{\rm p}R_{\rm S}^3/3$ becomes comparable with 
the ejecta mass $M_{\rm ej}$, where $n_{\rm ISM}$ and $m_{\rm p}$ are 
the number density of the interstellar medium (ISM) and the proton mass, 
respectively.
The free expansion velocity, $u_{\rm free}=R_{\rm S}/t_{\rm S}$, 
is obtained from $E_{\rm SN}=M_{\rm ej}u_{\rm free}^2/2$, 
where $E_{\rm SN}$ is the explosion energy of a supernova.
Then, $R_{\rm S}$ and $t_{\rm S}$ are represented by
\begin{eqnarray}
\label{eq:rs}
R_{\rm S}&=& 2.13~{\rm pc}~\left(\frac{M_{\rm ej}}{1M_{\odot}}\right)^{\frac{1}{3}} \left(\frac{n_{\rm ISM}}{1~{\rm cm}^{-3}}\right)^{-\frac{1}{3}}~~, \\
t_{\rm S}&=& 209~{\rm yr}~\left(\frac{E_{\rm SN}}{10^{51}~{\rm erg}}\right)^{-\frac{1}{2}} \left(\frac{M_{\rm ej}}{1M_{\odot}}\right)^{\frac{5}{6}} \left(\frac{n_{\rm ISM}}{1~{\rm cm}^{-3}}\right)^{-\frac{1}{3}}~~.
\label{eq:ts}
\end{eqnarray}
For simplicity, we assume $R_{\rm S}=2~{\rm pc}$ and $t_{\rm S}=200~{\rm yr}$ in this paper.

\subsection{Relevant timescales}
The maximum energy of accelerated particles is limited by a finite SNR
age, their cooling, or escape. Hence it is obtained by comparisons of timescales,
which are given as functions of a CR energy, $E$, and the SNR age, $t$,
(see Table~\ref{table1}).
The acceleration time of DSA, $t_{\rm acc}(E,t)$, is represented by
\begin{equation}
t_{\rm acc}(E,t) = \eta_{\rm acc}\frac{D(E,t)}{u_{\rm sh}(t)^2}~~,
\label{eq:tacc1}
\end{equation}
where 
$D(E,t)$ is the diffusion coefficient around the shock, and
$\eta_{\rm acc}\approx10$ is a numerical factor which depends on
the shock compression ratio and the spatial dependence of $D(E,t)$ \citep{drury83}.
First, we assume the Bohm-type diffusion of relativistic particles 
although the diffusion coefficient is still unclear when the magnetic
field is strongly amplified \citep[e.g.][]{reville08}.
The Bohm-type diffusion means that the mean free path of CRs is 
proportional to the gyroradius of CRs, where the constant of proportionality, 
$\eta_{\rm g}(t)$, is the gyrofactor and $\eta_{\rm g}=1$ for the Bohm limit. 
Then, the diffusion coefficient of CRs is represented by
\begin{equation}
D(E,t) = \eta_{\rm g}(t)\frac{cE}{3eB(t)}~~,
\label{eq:diffc}
\end{equation}
where $c$, $E$, $e$, and $B(t)$ are 
the velocity of light, the energy of CRs, the elementary
charge and the magnetic field strength in the upstream region, respectively.
In section~\ref{sec:magneticfield}, we will give explicit forms
of time dependence of $B(t)$ and $\eta_{\rm g}(t)$ 
[see Eqs.~(\ref{eq:b}) and (\ref{eq:etag})]. 
Then, the acceleration time can be expressed by
\begin{equation}
t_{\rm acc}(E,t) = \eta_{\rm acc}\eta_{\rm g}(t)\frac{cE}{3eB(t)u_{\rm sh}(t)^2}~~.
\label{eq:tacc2}
\end{equation}

The escape time due to diffusion, $t_{\rm esc}(E,t)$, is written by
\begin{equation}
t_{\rm esc}(E,t) = \eta_{\rm esc}\frac{R_{\rm sh}(t)^2}{D(E,t)}
=\frac{\eta_{\rm esc}}{\eta_{\rm g}(t)}\frac{3eB(t)R_{\rm sh}(t)^2}{cE}~~.
\label{eq:stesc}
\end{equation}
where $\eta_{\rm esc}(<1)$ is a numerical factor.

The energy loss of CR protons above $1~{\rm GeV}$ is due to the pion 
production by inelastic collisions, and the cooling time of CR protons, 
$t_{\rm cool,p}(t)$, is represented by
\begin{equation}
t_{\rm cool,p}(t) \approx \frac{1}{0.5n \sigma_{\rm pp}c}~~,
\label{eq:tcoolp}
\end{equation}
where $n$ and $\sigma_{\rm pp}\approx 3\times10^{-26}~{\rm cm}^2$ 
are the number density ($\sim 4n_{\rm ISM}$ in the downstream region) and the cross section of the nuclear interaction, 
respectively.
A factor of 0.5 is the inelasticity of the nuclear interaction.
Cooling of CR electrons above $1~{\rm GeV}$ is due to synchrotron 
emission, inverse Compton scattering, or bremsstrahlung emission.
For typical SNRs, the energy loss is dominated by synchrotron emission 
in the downstream region when the maximum energy is limited by cooling.
The cooling time of CR electrons due to synchrotron emission in the 
downstream region, $t_{\rm cool,e}(E,t)$, is represented by
\begin{equation}
t_{\rm cool,e}(E,t) = \frac{9m_{\rm e}^4c^7}{4e^4B_{\rm d}(t)^2E}~~,
\label{eq:tcoolsyn}
\end{equation}
where $m_{\rm e}$ is the electron mass and $B_{\rm d}(t)$ is the magnetic 
field strength in the downstream region.
In this paper, we assume $B_{\rm d}(t)=4B(t)$ because of the shock 
compression.
\subsection{Maximum energy of CR protons}
For CR protons, cooling is usually not important 
for a determination of the maximum energy.
One can obtain the age-limited maximum energy, $E_{\rm m,age}(t)$, 
from the condition $t_{\rm acc}(E,t)=t$ as
\begin{eqnarray}
E_{\rm m,age}(t)
=\frac{3eB(t)R_{\rm S}^2}{\eta_{\rm acc}\eta_{\rm g}(t)ct_{\rm S}}
\times \left\{ \begin{array}{ll}
\left(\frac{t}{t_{\rm S}}\right)          & ~(~t \leq t_{\rm S}~) \\
\left(\frac{t}{t_{\rm S}}\right)^{-\frac{1}{5}}& ~(~t_{\rm S}\leq t~) \\
\end{array} \right. ~~,
\label{eq:emage1}
\end{eqnarray}
while the escape-limited maximum energy, $E_{\rm m,esc}(t)$, is obtained 
from the condition $t_{\rm acc}(E,t)=t_{\rm esc}(E,t)$ as
\begin{equation}
E_{\rm m,esc}(t) = \sqrt{\eta_{\rm esc}\eta_{\rm acc}}E_{\rm m,age}(t)~~.
\label{eq:emesc1}
\end{equation}
Note that $E_{\rm m,age}(t)$ is comparable to $E_{\rm m,esc}(t)$ 
because $\sqrt{\eta_{\rm esc}\eta_{\rm acc}}$ is on the order of 
unity \citep{drury11}.
In this paper, we assume $E_{\rm m,age}(t)=E_{\rm m,esc}(t)$
for simplicity (that is, $\sqrt{\eta_{\rm esc}\eta_{\rm acc}}=1$). 
If the maximum energy is limited by escape, CRs above 
$E_{\rm m,esc}(t)$ leave the shock front upstream.
However, they are caught up by the shock during the free 
expansion phase because the diffusion length is proportional to 
$t^{1/2}$ but the shock radius is proportional to $t$.
Therefore, the maximum energy is limited by the SNR age during the free expansion phase.
On the other hand, CRs can leave the shock front during the Sedov phase 
because $R_{\rm sh}\propto t^{2/5}$, which is slower than diffusion.
Therefore, the maximum energy is limited by escape during the Sedov phase.

As shown in Equations (\ref{eq:emage1}) and (\ref{eq:emesc1}), 
the maximum energy depends on the evolution of the magnetic field 
around the shock \citep[e.g.][]{ptuskin03}.
Although the magnetic field amplification around the shock is studied 
by linear analyses \citep{bell04,reville07,ohira09b,ohira10,bykov11,schure11} and 
numerical simulations \citep{lucek00,giacalone07,
niemiec08,riquelme09,inoue09,ohira09a,vladimirov09,gargate10}, 
the saturation of the magnetic-field amplification and the diffusion 
coefficient have not been understood yet in detail.
Here we use a phenomenological approach based on the 
assumption that young SNRs are responsible for observed CRs 
below the knee \citep{gabici09}. 
The maximum energy of CR protons, $E_{\rm m,p}$(t), is expected to 
increase up to the knee energy $E_{\rm knee}=10^{15.5}~{\rm eV}$ 
until the beginning of the Sedov phase $t_{\rm S}$ and decreases 
from that epoch. 
We may assume a functional form of $E_{\rm m,p}(t)$ to be
\begin{eqnarray}
E_{\rm m,p}(t)=E_{\rm knee}\times \left\{ \begin{array}{ll}
\left(\frac{t}{t_{\rm S}}\right)                & ~(~t \leq t_{\rm S}~) \\
\left(\frac{t}{t_{\rm S}}\right)^{-\alpha} & ~(~t_{\rm S}\leq t~) \\
\end{array} \right. ~~,
\label{eq:emaxp}
\end{eqnarray}
where $\alpha$ is a parameter to describe the evolution of the maximum 
energy during the Sedov phase. 
In this paper, we assume that $E_{\rm m,p}=1~{\rm GeV}$ at the end of the Sedov phase 
$t=10^{2.5}t_{\rm Sedov}$ to reproduce Galactic CRs, then $\alpha=2.6$.
Hereafter, we adopt $\alpha=2.6$.
This assumption is consistent with the following fact.
So far, about 300 radio SNRs have been observed in a part of our 
Galaxy \citep{case98}.
From \citet{case98}, we expect that there are about 
$5 \times 300$ radio SNRs in our Galaxy.
Assuming the supernova rate of $0.03~{\rm yr}^{-1}$, 
the lifetime of radio SNRs is about $5\times 10^{4}~{\rm yr}$ 
which is comparable to the end time of the Sedov phase.
Radio synchrotron photons are emitted by electrons with energies of a few GeV, 
so that we can expect that the lifetime of radio SNRs 
corresponds to the escape time of $1~{\rm GeV}$ particles. 
Therefore, the assumption, $E_{\rm m,p}(t=10^{2.5}t_{\rm Sedov})=1~{\rm GeV}$, is reasonable.
The transition time from the Sedov phase to the radiative phase is somewhat ambiguous.
\citet{truelove99} and \citet{petruk05} thought the end of the Sedov phase is 
$1.2\times10^4~{\rm yr}$ and $3\times10^4~{\rm yr}$, respectively. 
In these cases, $\alpha$ becomes 3.66 and 2.99, respectively.
Future direct observations of CR electrons, AMS-02, CALET, CTA and 
LHAASO will be able to provide useful informations about $\alpha$ 
\citep{kawanaka11,thoudam12}.

Assuming  Equation~(\ref{eq:emaxp}), the age-limited maximum energy 
$E_{\rm m,age}$ for $t\leq t_{\rm S}$ can be expressed by
\begin{equation}
E_{\rm m,age}(t) = E_{\rm knee} \left(\frac{t}{t_{\rm S}}\right)~~,
\label{eq:emage2}
\end{equation}
while for $t\geq t_{\rm S}$, the escaped-limited maximum energy $E_{\rm m,esc}$
can be expressed by 
\begin{equation}
E_{\rm m,esc}(t) = E_{\rm knee} \left(\frac{t}{t_{\rm S}}\right)^{-\alpha}~~.
\label{eq:emesc2}
\end{equation}
Moreover, from Equation~(\ref{eq:emesc2}), the time when CRs with an energy of $E$ 
escape from the SNR, $T_{\rm esc}(E)$, can be expressed by
\begin{equation}
T_{\rm esc}(E) = t_{\rm S} \left(\frac{E}{E_{\rm knee}}\right)^{-\frac{1}{\alpha}}~~.
\label{eq:tesc}
\end{equation}
From Equations~(\ref{eq:rsh}) and (\ref{eq:tesc}), the escape radius of CRs, 
that is, the SNR radius at which CRs with an energy of $E$ escape, 
can be expressed by
\begin{equation}
R_{\rm esc}(E) = R_{\rm S} \left(\frac{E}{E_{\rm knee}}\right)^{-\frac{2}{5\alpha}}~~.
\label{eq:resc}
\end{equation}
\subsection{Magnetic field}
\label{sec:magneticfield}

Equations (\ref{eq:emage1}), (\ref{eq:emesc1}), and (\ref{eq:emaxp})
determine the evolution of $B(t)/\eta_{\rm g}(t)$.
During the free-expansion phase, both the upstream magnetic field and the 
gyrofactor are constant with time: $B(t<t_{\rm S})=B_{\rm free}$ and 
$\eta_g(t<t_{\rm S})=\eta_{\rm g,free}$, because these quantities may depend on the 
shock velocity which is constant in this phase.
For the Sedov phase, we assume that the upstream magnetic field strength
is $B \propto t^{-\alpha_{\rm B}}$ as long as it
is larger than the value of the ISM, 
$B_{\rm ISM}$, and that after the end time 
of the magnetic field amplification $t_B$ at which
$B(t_{\rm B})=B_{\rm ISM}$, $B$ is equal to $B_{\rm ISM}$.
Then, the upstream magnetic field, $B(t)$, and the gyrofactor, $\eta_{\rm g}(t)$, 
are given by
\begin{eqnarray}
B(t)=\left\{ \begin{array}{ll}
B_{\rm free}          & ~(~t \leq t_{\rm S}~) \\
B_{\rm free}\left(\frac{t}{t_{\rm S}}\right)^{-\alpha_{\rm B}}& ~(~t_{\rm S} \leq  t \leq  t_{\rm B}~) \\
B_{\rm ISM} & ~(~t_{\rm B} \leq t ~) \\
\end{array} \right. ~~,
\label{eq:b}
\end{eqnarray}
and
\begin{eqnarray}
\eta_{\rm g}(t)=\eta_{\rm g,free} \times \left\{ \begin{array}{ll}
1                                                                                                 & ~(~t \leq  t_{\rm S}~) \\
\left(\frac{t}{t_{\rm S}}\right)^{\alpha-\alpha_{\rm B}-\frac{1}{5}} & ~(~t_{\rm S} \leq t \leq  t_{\rm B}~) \\
\left(\frac{t_{\rm B}}{t_{\rm S}}\right)^{-\alpha_{\rm B}}\left(\frac{t}{t_{\rm S}}\right)^{\alpha-\frac{1}{5}}                            & ~(~t_{\rm B} \leq t ~) \\
\end{array} \right. ~~,
\label{eq:etag}
\end{eqnarray}
where $B_{\rm free}$ is the amplified magnetic field during the free expansion phase and given by
\begin{eqnarray}
B_{\rm free} &=& \frac{\eta_{\rm g,free}\eta_{\rm acc}ct_{\rm S}E_{\rm knee}}{3eR_{\rm S}^2} \nonumber \\
&=& 174~{\rm \mu G}\left(\frac{\eta_{\rm g,free}}{1}\right)\left(\frac{\eta_{\rm acc}}{10}\right)  \nonumber \\
&&\times \left(\frac{E_{\rm knee}}{10^{15.5}~{\rm eV}}\right) \left(\frac{t_{\rm S}}{200~{\rm yr}}\right) \left(\frac{R_{\rm S}}{2~{\rm pc}}\right)^{-2}~~,
\label{eq:bfree}
\end{eqnarray}
and $\eta_{\rm g,free}\approx1$ is the gyrofactor during the 
free expansion phase, and the end time of the magnetic field amplification 
$t_{B}$ is given by
\begin{equation}
t_{\rm B} = t_{\rm S} \left(\frac{B_{\rm free}}{B_{\rm ISM}}\right)^{\frac{1}{\alpha_{\rm B}}} ~~.
\label{eq:tb1}
\end{equation}
The downstream magnetic field is $B_{\rm d}=4B_{\rm free}=697~{\rm \mu G}$ during the free expansion phase.
In this paper, we consider the following three evolution models of $B(t)$ for $t_{\rm S}<t<t_{\rm B}$, 
\begin{eqnarray}
\alpha_{\rm B}=\left\{ \begin{array}{ll}
\alpha-\frac{1}{5}  & ~(~{\rm for}~~ \eta_{\rm g}=\eta_{\rm g,free}~) \\
\frac{9}{10}           & ~(~{\rm for}~~ B^2 \propto u_{\rm sh}^3~) \\
\frac{3}{5}             & ~(~{\rm for}~~ B^2 \propto u_{\rm sh}^2~) \\
\end{array} \right. ~~.
\label{eq:alphab}
\end{eqnarray}
The first model originates from the assumption that the gyrofactor,
$\eta_{\rm g}$, is  constant  during $B>B_{\rm ISM}$. 
In this case, $B^2\propto t^{-4.8}\propto u_{\rm sh}^8$ for $\alpha=2.6$.
The second model is proposed by \citet{bell04}.
The third model originates from the assumption that the pressure of the amplified magnetic field is proportional to the shock ram pressure \citep[e.g.][]{voelk05}.
Observations of the velocity dependence of the magnetic field can be found 
in \citet{vink08} that suggests $B^2 \propto u_{\rm sh}^3$.

From Equations~(\ref{eq:bfree}) - (\ref{eq:alphab}), 
the end time of the magnetic field amplification, $t_{\rm B}$, 
are given by
\begin{eqnarray}
t_{\rm B}=\left\{ \begin{array}{ll}
1.09\times 10^3~{\rm yr} & ~(~{\rm for}~~ \eta_{\rm g}=\eta_{\rm g,free}~) \\
1.82\times 10^4~{\rm yr} & ~(~{\rm for}~~ B^2 \propto u_{\rm sh}^3~) \\
1.74\times 10^5~{\rm yr} & ~(~{\rm for}~~ B^2 \propto u_{\rm sh}^2~) \\
\end{array} \right. ~~,
\label{eq:tb2}
\end{eqnarray}
where we assume $\alpha=2.6$ and $B_{\rm ISM}=3~{\rm \mu G}$. 
Note that we do not discuss the density dependence of the magnetic field in this paper.
Furthermore, there is no guarantee that 
the magnetic field evolution is expressed by a power law form \citep{ptuskin03,yan12}.
However, the Galactic CR spectrum is approximated by a single power law below 
the knee energy, that is, CR observations suggest absence of characteristic 
scale.
Moreover, the power-law behavior of the gyrofactor, $\eta_{\rm g}\propto t^{\alpha-1/5}$, is expected by \citet{ptuskin03,yan12}. 
Therefore, we assume that all evolutions have power law dependences.
We believe that this approximation is useful to extract essential features.
\subsection{Maximum energy of CR electrons}
\begin{table*}
\centering
\begin{minipage}{180mm}
\caption{Quantities and parameters appearing in Section 2}
\label{table1}
\begin{tabular}{lll}
\hline
 & Characteristic times  & \\
\hline
$t_{\rm acc}(E,t)$    & acceleration time              & Eqs.~(\ref{eq:tacc1}),~(\ref{eq:tacc2}) \\
$t_{\rm esc}(E,t)$    & escape time due to diffusion   & Eq.~(\ref{eq:stesc}) \\
$t_{\rm cool,p}(t)$   & cooling time of CR protons     & Eq.~(\ref{eq:tcoolp}) \\
$t_{\rm cool,e}(E,t)$ & cooling time of CR electrons   & Eq.~(\ref{eq:tcoolsyn}) \\
$t_{\rm c}$        & start time of cooling & Eq.~(\ref{eq:tc}) \\
$t_{\rm S}=200$~yr & start time of Sedov phase & Eqs.~(\ref{eq:rsh}),~(\ref{eq:ts}) \\
$t_{\rm e}$        & start time of escape of CR electrons & Eq.~(\ref{eq:te}) \\
$t_{\rm B}$        & end time of magnetic field amplification & Eqs.~(\ref{eq:tb1}),~(\ref{eq:tb2}) \\
$T_{\rm esc}(E)$ & time when CRs with an energy of $E$ escape from the SNR   & Eq.~(\ref{eq:tesc}) \\
\hline
 & Characteristic energies & \\
\hline
$E_{\rm m, age}(t) $  & given by $t_{\rm acc}(E,t)=t$                 & Eqs.~(\ref{eq:emage1}),~(\ref{eq:emage2}) \\
$E_{\rm m, esc}(t) $  & given by $t_{\rm acc}(E,t)=t_{\rm esc}(E,t)$  & Eqs.~(\ref{eq:emesc1}),~(\ref{eq:emesc2}) \\
$E_{\rm m, cool}(t)$  & given by $t_{\rm acc}(E,t)=t_{\rm cool}(E,t)$ & Eq.~(\ref{eq:emaxcool}) \\
$E_{\rm b}(t)$        & given by $t_{\rm cool}(E,t)=t$                & Eq.~(\ref{eq:eb}) \\
$E_{\rm m,p}(t)$      & maximum energy of CR protons                  & Eq.~(\ref{eq:emaxp}) \\
$E_{\rm m,e}(t)$      & maximum energy of CR electrons ($\min \{E_{\rm m, age}(t), E_{\rm m, esc}(t), E_{\rm m, cool}(t) \}$)               & Eqs.~(\ref{eq:eme1}),~(\ref{eq:emaxe}),~(\ref{eq:emaxe2}) \\
$E_{\rm m,S}=E_{\rm m,cool}(t<t_{\rm S})$ & cooling-limited maximum energy during the free expansion phase  & Eq.~(\ref{eq:ems}) \\
$E_{\rm b,S}=E_{\rm b}(t_{\rm S})$   & break energy at the beginning of the Sedov phase  & Eq.~(\ref{eq:ebs}) \\
\hline
 & Other quantities evolving with time & \\
\hline
$R_{\rm sh}(t)$      & SNR shock radius                  & Eq.~(\ref{eq:rsh})  \\
$u_{\rm sh}(t)$      & shock velocity                    & Eq.~(\ref{eq:ush})  \\
$D(E,t)$                   & diffusion coefficient around shocks                          & Eqs.~(\ref{eq:tacc1}),~(\ref{eq:diffc}) \\
$B(t)$               & upstream magnetic field           & Eqs.~(\ref{eq:diffc}),~(\ref{eq:b}) \\
$B_{\rm d}(t)=4B(t)$ & downstream magnetic field         & \\
$\eta_{\rm g}(t)$    & gyrofactor                        & Eqs.~(\ref{eq:diffc}), (\ref{eq:etag}) \\
\hline
 & Others & \\
\hline
$t$     & SNR age           &   \\
$E$     & CR energy         & \\
$e$     & electron charge   & \\
$c$     & velocity of light & \\
$R_{\rm S} =2$~pc          & SNR radius at $t_{\rm S}$                                 & Eqs.~(\ref{eq:rsh}),~(\ref{eq:rs}) \\
$R_{\rm esc}(E)$      & escape radius of CRs                  & Eq.~(\ref{eq:resc})  \\
$D_{\rm ISM}(E)$           & diffusion coefficient in ISM                              & Eq.~(\ref{eq:dism}) \\
$E_{\rm knee}=10^{15.5}$eV & maximum energy of CR protons at $t_{\rm S}$               & Eq.~(\ref{eq:emaxp}) \\
$B_{\rm ISM}=3~\mu$G       & ISM magnetic field strength                               & Eqs.~(\ref{eq:b}),~(\ref{eq:tb1}) \\
$B_{\rm free}=174~\mu$G    & amplified field strength during the free expansion phase  & Eqs.~(\ref{eq:b}),~(\ref{eq:bfree}) \\
$\eta_{\rm g,free}=1$      & gyrofactor during the free expansion phase                & Eq.~(\ref{eq:etag}) \\
$\eta_{\rm acc}=10$        & numerical factor for the acceleration time                & Eq.~(\ref{eq:tacc1}) \\
$\eta_{\rm esc}=0.1$       & numerical factor for the escape time                      & Eq.~(\ref{eq:stesc}) \\
$n_{\rm ISM}=1$~cm$^{-3}$            & ISM density                                               & Eq.~(\ref{eq:rs})\\
$\alpha_{\rm B}$           & temporal decay index of the magnetic field ($B\propto t^{-\alpha_{\rm B}}$) & Eqs.~(\ref{eq:b}),~(\ref{eq:alphab}) \\
$\alpha=2.6$               & temporal decay index of the maximum energy of CR protons ($E_{\rm m,p}\propto t^{-\alpha}$) & Eqs.~(\ref{eq:emaxp}),~(\ref{eq:nesc}) \\
$\beta=0.6~(t>t_{\rm S})$           & temporal index of the number of CRs electrons inside SNRs (${\rm d}N/{\rm d}E\propto t^{\beta}E^{-s}$)& Eq.~(\ref{eq:nesc}) \\
$s=2.0$                    & spectral index of CR electrons inside SNRs (${\rm d}N/{\rm d}E\propto t^{\beta}E^{-s}$)                  & Eq.~(\ref{eq:nesc})\\
$A$           & normalization of the spectrum of runaway CR electrons (${\rm d}N_{\rm esc}/{\rm d}E =AE^{-(s+\beta/\alpha)}$)& Eq.~(\ref{eq:nesc}) \\
\hline
\end{tabular}
\end{minipage}
\end{table*}
\begin{figure}
\begin{center}
\includegraphics[width=80mm]{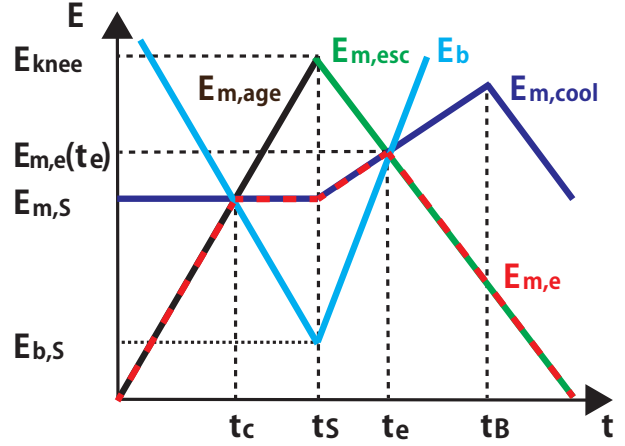}
\end{center}
\caption{Schematic picture of evolutions of the maximum energy and 
the break energy of CR electrons for $\eta_{\rm g}=\eta_{\rm g,free}$ 
and $t_{\rm e}<t_{\rm B}$. 
The black, blue, and green lines show the age-limited, the cooling-limited, 
and the escape-limited maximum energy, respectively.
The red dashed line shows the maximum energy of CR electrons.
The cyan line shows the break energy owing to synchrotron cooling.
$t_{\rm S}$, $t_{\rm e}$, and $t_{\rm B}$ are the SNR age at the beginning 
of the Sedov phase, the start time of escape of CR electrons}, and the end time of the 
magnetic field amplification, respectively.
$E_{\rm m,e}(t_{\rm e})$ is the maximum energy of runaway CR electrons.
\label{fig:1}
\end{figure}
\begin{figure}
\begin{center}
\includegraphics[width=80mm]{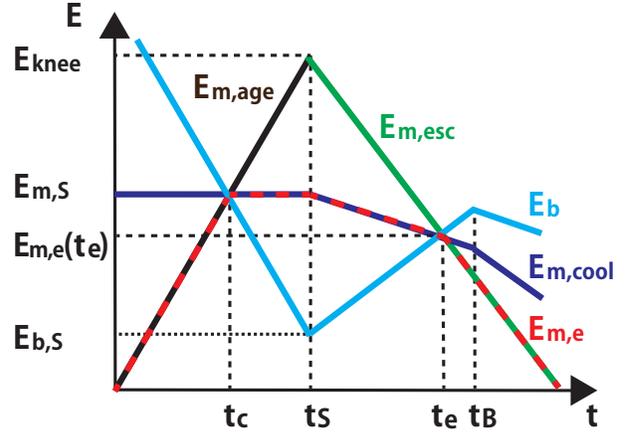}
\end{center}
\caption{The same as Figure~\ref{fig:1}, but for $B^2\propto u_{\rm sh}^2~{\rm or}~u_{\rm sh}^3$.}
\label{fig:2}
\end{figure}
The maximum energy of CR electrons is limited by the SNR age 
at early stages. 
As the maximum energy increases, synchrotron cooling becomes 
significant and limits the maximum energy.
From the condition $t_{\rm acc}(E,t)=t_{\rm cool,e}(E,t)$ and Equations
(\ref{eq:b}) and (\ref{eq:etag}), we obtain the cooling-limited
maximum energy, $E_{\rm m,cool}(t)$, as
\begin{eqnarray}
E_{\rm m,cool}(t)=E_{\rm m,S} \times\left\{ \begin{array}{ll}
1          & (t \leq t_{\rm S}) \\
\left(\frac{t}{t_{\rm S}}\right)^{\frac{2\alpha_{\rm B}-\alpha-1}{2}}& (t_{\rm S} \leq  t\leq t_{\rm B} ) \\
\left(\frac{t_{\rm B}}{t_{\rm S}}\right)^{\alpha_{\rm B}}\left(\frac{t}{t_{\rm S}}\right)^{-\frac{\alpha+1}{2}}& (t_{\rm B} \leq t)\\
\end{array} \right. ~~,
\label{eq:emaxcool}
\end{eqnarray}
where $E_{\rm m,S}$ is the cooling-limited maximum energy during the
free expansion phase ($t<t_{\rm S}$) and given by
\begin{eqnarray}
E_{\rm m,S} &=& \frac{9m_{\rm e}^2c^{5/2}R_{\rm S}^2}{8\eta_{\rm g,free}\eta_{\rm acc}et_{\rm S}^{3/2}E_{\rm knee}^{1/2}} \nonumber \\
&=&2.02\times10^{13}~{\rm eV} \left(\frac{\eta_{\rm g,free}}{1}\right)^{-1}\left(\frac{\eta_{\rm acc}}{10}\right)^{-1} \nonumber \\
&& \times \left(\frac{E_{\rm knee}}{10^{15.5}~{\rm eV}}\right)^{-\frac{1}{2}} \left(\frac{t_{\rm S}}{200~{\rm yr}}\right)^{-\frac{3}{2}} \left(\frac{R_{\rm S}}{2~{\rm pc}}\right)^2~~. 
\label{eq:ems}
\end{eqnarray}
$E_{\rm m,S}$ can be obtained from only physical values at the beginning 
of the Sedov phase.
In this paper, SNRs are characterized by only $t_{\rm S}$ and $R_{\rm S}$.
Therefore, our results during $t>t_{\rm S}$ do not depend on any assumptions 
during $t<t_{\rm S}$ in our formalism.

The evolution of the maximum energy of CR electrons, $E_{\rm m,e}(t)$, 
is given by 
\begin{equation}
E_{\rm m,e}(t) = \min \left \{E_{\rm m,age}(t), E_{\rm m,cool}(t), E_{\rm m,esc}(t)\right \}~~. 
\label{eq:eme1}
\end{equation}
Conditions $E_{\rm m,age}(t_{\rm c})=E_{\rm m,cool}(t_{\rm c})$ 
for the free expansion phase ($t<t_{\rm S}$) and 
$E_{\rm m,cool}(t_{\rm e})= E_{\rm m,esc}(t_{\rm e})$ 
for the Sedov phase ($t_{\rm S}<t$)
provide two transition times, $t_{\rm c}$ and $t_{\rm e}$, respectively 
(see Figures~(\ref{fig:1}) and (\ref{fig:2})).

\subsubsection{The case of $t_{\rm e}<t_{\rm B}$}
\label{sec:2.5.1}
We find $t_{\rm e}<t_{\rm B}$ for typical SNRs with $\alpha\approx2.6$.
In this case,
 the maximum energy of CR electrons, $E_{\rm m,e}$, can be expressed by
\begin{eqnarray}
E_{\rm m,e}(t)=\left\{ \begin{array}{ll}
E_{\rm knee} \left(\frac{t}{t_{\rm S}}\right) & (t \leq t_{\rm c}) \\
E_{\rm m,S}& (t_{\rm c} \leq t\leq t_{\rm S} ) \\
E_{\rm m,S}\left(\frac{t}{t_{\rm S}}\right)^{\frac{2\alpha_{\rm B}-\alpha-1}{2}}& (t_{\rm S} \leq t<t_{\rm e})\\
E_{\rm knee}\left(\frac{t}{t_{\rm S}}\right)^{-\alpha}&(t_{\rm e} \leq t)\\
\end{array} \right. ~~,
\label{eq:emaxe}
\end{eqnarray}
where $t_{\rm c}$ is given by
\begin{eqnarray}
t_{\rm c} &=& t_{\rm S} \frac{E_{\rm m,S}}{E_{\rm knee}} \nonumber \\
&=&1.28~{\rm yr} \left(\frac{\eta_{\rm g,free}}{1}\right)^{-1} \left(\frac{\eta_{\rm acc}}{10}\right)^{-1}  \nonumber \\
&& \times \left(\frac{E_{\rm knee}}{10^{15.5}~{\rm eV}}\right)^{-\frac{3}{2}} \left(\frac{t_{\rm S}}{200~{\rm yr}}\right)^{-\frac{1}{2}} \left(\frac{R_{\rm S}}{2~{\rm pc}}\right)^2~~,
\label{eq:tc}
\end{eqnarray}
and  $t_{\rm e}$ is the start time of escape of CR electrons and given by
\begin{eqnarray}
t_{\rm e}&=&t_{\rm S}\left(\frac{E_{\rm m,S}}{E_{\rm knee}}\right)^{-\frac{2}{\alpha+2\alpha_{\rm B}-1}}  \nonumber \\
&=& \left(\frac{9m_{\rm e}^2c^{5/2}R_{\rm S}^2}{8\eta_{\rm g,free}\eta_{\rm acc}eE_{\rm knee}^{3/2}} \right)^{-\frac{2}{\alpha+2\alpha_{\rm B}-1}}t_{\rm S}^{\frac{\alpha+2\alpha_{\rm B}+2}{\alpha+2\alpha_{\rm B}-1}} \nonumber \\
&=&\left\{ \begin{array}{ll}
9.70\times 10^2~{\rm yr}& ~(~{\rm for}~~ \eta_{\rm g}=\eta_{\rm g,free}~) \\
3.91\times 10^3~{\rm yr}& ~(~{\rm for}~~ B^2 \propto u_{\rm sh}^3~) \\
7.39\times 10^3~{\rm yr}& ~(~{\rm for}~~ B^2 \propto u_{\rm sh}^2~) \\
\end{array} \right. ~~,
\label{eq:te}
\end{eqnarray}
where we assume $\alpha=2.6$, Equation~(\ref{eq:alphab}) for 
$\alpha_{\rm B}$ and normalizations of Equation~(\ref{eq:ems}).

In Figure~\ref{fig:1}, we show the schematic picture of the evolution of the 
maximum energy as a function of time (red dashed line) for 
$\eta_{\rm g}=\eta_{\rm g,free}$ (see also Table~1).
In Figure~\ref{fig:2}, we show the same figure as Figure~\ref{fig:1}, 
but for $B^2\propto u_{\rm sh}^2~{\rm or}~u_{\rm sh}^3$.
Initially, the maximum energy increases linearly with time ($t<t_{\rm c}$), 
then the maximum energy is constant until the beginning of the Sedov 
phase ($t_{\rm c}<t<t_{\rm S}$). 
During $t_{\rm S}<t<t_{\rm e}$, the maximum energy of CR electrons 
increases with time 
for $\eta_{\rm g}=\eta_{\rm g,free}$ (Figure~\ref{fig:1}) and decreases 
for $B^2\propto u_{\rm sh}^2~{\rm or}~u_{\rm sh}^3$ (Figure~\ref{fig:2}).
Finally, CR electrons start to escape from the SNR at $t=t_{\rm e}$. 
The maximum energy of runaway CR electrons is given by
\begin{eqnarray}
E_{\rm m,e}(t_{\rm e}) &=& E_{\rm knee}\left(\frac{E_{\rm m,S}}{E_{\rm knee}}\right)^{\frac{2\alpha}{\alpha+2\alpha_{\rm B}-1}} \nonumber \\
&=& \left(\frac{9m_{\rm e}^2c^{5/2}R_{\rm S}^2}{8\eta_{\rm g,free}\eta_{\rm acc}et_{\rm S}^{3/2}} \right)^{\frac{2\alpha}{\alpha+2\alpha_{\rm B}-1}}E_{\rm knee}^{-\frac{2\alpha-2\alpha_{\rm B}+1}{\alpha+2\alpha_{\rm B}-1}} \nonumber \\
&=&\left\{ \begin{array}{ll}
5.21 \times 10^{13}~{\rm eV}& ~(~{\rm for}~~ \eta_{\rm g}=\eta_{\rm g,free}~) \\
1.39 \times 10^{12}~{\rm eV}& ~(~{\rm for}~~ B^2 \propto u_{\rm sh}^3~) \\
2.66 \times 10^{11}~{\rm eV}& ~(~{\rm for}~~ B^2 \propto u_{\rm sh}^2~) \\
\end{array} \right. ~~,
\label{eq:eescmax}
\end{eqnarray}
where we assume $\alpha=2.6$, Equation~(\ref{eq:alphab}) for 
$\alpha_{\rm B}$ and normalizations of Equation~(\ref{eq:ems}).
Therefore, SNRs can produce the Galactic CR electrons up to about 
$0.3-50~{\rm TeV}$  which depends on the evolution of the magnetic field.
Interestingly, the maximum energy of runaway CR electrons becomes smaller as the maximum energy of CR protons ($E_{\rm knee}$) becomes larger.

\subsubsection{The case of $t_{\rm e}>t_{\rm B}$}
\label{sec:2.5.2}
If $\alpha$ is smaller, $t_{\rm e}$ can be larger than $t_{\rm B}$.
In this case, $E_{\rm m,e}$ can be expressed by 
\begin{eqnarray}
E_{\rm m,e}(t)=\left\{ \begin{array}{ll}
E_{\rm knee} \left(\frac{t}{t_{\rm S}}\right) & (t \leq t_{\rm c}) \\
E_{\rm m,S}& (t_{\rm c} \leq t\leq t_{\rm S} ) \\
E_{\rm m,S}\left(\frac{t}{t_{\rm S}}\right)^{\frac{2\alpha_{\rm B}-\alpha-1}{2}}& (t_{\rm S} \leq t<t_{\rm B})\\
E_{\rm m,S}\left(\frac{t_B}{t_{\rm S}}\right)^{\alpha_{\rm B}} \left(\frac{t}{t_{\rm S}}\right)^{-\frac{\alpha+1}{2}}& (t_{\rm B} \leq t<t_{\rm e})\\
E_{\rm knee}\left(\frac{t}{t_{\rm S}}\right)^{-\alpha}&(t_{\rm e} \leq t)\\
\end{array} \right. ~~,
\label{eq:emaxe2}
\end{eqnarray}
where we assume $\alpha>1$, otherwise $E_{\rm m,cool}$ is always
smaller than $E_{\rm m,esc}$, that is, CR electrons can not escape from 
the SNR until the SNR shock disappears.
The start time of escape of CR electrons, $t_{\rm e}$, for $t_{\rm e}>t_{\rm B}$ 
is obtained by the condition $E_{\rm m,esc}(t)=E_{\rm m,cool}(t)$ for $t_{\rm B}<t$,
\begin{equation}
t_{\rm e} = t_{\rm S} \left( \frac{E_{\rm m,S}}{E_{\rm knee}} \right)^{ -\frac{2}{\alpha-1}} \left(\frac{t_{\rm B}}{t_{\rm S}}\right)^{ -\frac{2\alpha_{\rm B}}{\alpha-1} }~~.
\end{equation}
%
\subsection{Cooling break of CR electrons}
\label{sec:2.6}
The evolution of the CR spectrum inside SNRs is necessary to understand
those of runaway CRs \citep{ohiraetal10}.
CR electrons accumulate in SNRs with time,
and they cool down due to synchrotron radiation during the accumulation. 
Hence, the numbers of cooling CR electrons and non-cooling CR electrons 
are approximately given by $t_{\rm cool,e}(E) q(E)\propto E^{-1}q(E)$ and 
$t q(E)$, respectively, where $q(E)$ is an injection spectrum per unit time.
Therefore, the CR electron spectrum inside SNRs has a broken power 
law form (this is, so called, the cooling break).
The break energy, $E_{\rm b}(t)$, is obtained by $t_{\rm cool,e}(E_{\rm b},t)=t$.
From Equations (\ref{eq:tcoolsyn}) and (\ref{eq:b}), the break energy can be represented by
\begin{eqnarray}
E_{\rm b}(t)=E_{\rm b,S} \times \left\{ \begin{array}{ll}
\left(\frac{t}{t_{\rm S}}\right)^{-1} & (t\leq t_{\rm S}) \\
\left(\frac{t}{t_{\rm S}}\right)^{2\alpha_{\rm B}-1}& (t_{\rm S} \leq  t\leq t_{\rm B} ) \\
\left(\frac{t_{\rm B}}{t_{\rm S}}\right)^{2\alpha_{\rm B}} \left(\frac{t}{t_{\rm S}}\right)^{-1}& (t_{\rm B} \leq t)\\
\end{array} \right. ~~,
\label{eq:eb}
\end{eqnarray}
where $E_{\rm b,S}$ is the break energy at the beginning of the Sedov phase ($t=t_{\rm S}$) and given by
\begin{eqnarray}
E_{\rm b,S} &=& E_{\rm knee}\left(\frac{E_{\rm m,S}}{E_{\rm knee}}\right)^2\nonumber \\
&=&1.29\times10^{11}~{\rm eV} \left(\frac{\eta_{\rm g,free}}{1}\right)^{-2} \left(\frac{\eta_{\rm acc}}{10}\right)^{-2}  \nonumber \\
&& \times \left(\frac{E_{\rm knee}}{10^{15.5}~{\rm eV}}\right)^{-2} \left(\frac{t_{\rm S}}{200~{\rm yr}}\right)^{-3} \left(\frac{R_{\rm S}}{2~{\rm pc}}\right)^{4} ~~.
\label{eq:ebs}
\end{eqnarray}

In Figures~\ref{fig:1} and \ref{fig:2}, we show the evolution of the break energy 
$E_{\rm b}(t)$ as a function of time (Cyan line).
During $E_{\rm b}(t)<E_{\rm m,e}(t)$, the spectrum of CR electrons inside the SNR 
has the cooling break.
Interestingly, the break energy becomes the same as the maximum energy of CR 
electrons ($E_{\rm m,e}=E_{\rm b}$) at the start time of escape of CR electrons ($t=t_{\rm e}$).
The reason is as follows.
Characteristic energies, $E_{\rm m,age}(t)$, $E_{\rm m,esc}(t)$, $E_{\rm cool,e}(t)$, and
$E_{\rm b}(t)$ are obtained from the conditions 
$t_{\rm acc}(E,t)=t$, $t_{\rm esc}(E,t)=t_{\rm acc}(E,t)$, $t_{\rm acc}(E,t)=t_{\rm cool,e}(E,t)$, 
and $t_{\rm cool,e}(E,t)=t$, respectively (see also Table~\ref{table1}).
The SNR age is approximately the same as the escape 
time of just escaping CRs ($t \approx t_{\rm esc}$) because of 
$E_{\rm m,age}(t) \approx E_{\rm m,esc}(t)$ during the Sedov phase.
CR electrons start to escape when the cooling-limited maximum energy
becomes the same as the escape-limited maximum energy 
($E_{\rm cool,e}=E_{\rm m,esc}$), that is, the cooing time of CR electrons becomes
the same as the escape time ($t_{\rm cool,e} \approx t_{\rm esc}$).
Therefore, the SNR age is approximately the same as the cooling time 
$(t\approx t_{\rm cool,e})$ at $t_{\rm e}$.
This is the condition to derive the break energy.
Hence, all the characteristic energies have the same energy, $E_{\rm m,e}(t_{\rm e})$, at $t_{\rm e}$.

Namely, the spectrum of CR electrons inside the SNR has a cooling break 
before CR electrons escape ($t< t_{\rm e}$).
However, it does not appear while CR electrons escape ($t> t_{\rm e}$).
\subsection{Spectrum of runaway CR electrons}
As mentioned above, the spectrum of runaway CR electrons is unaffected by 
synchrotron cooling below $E_{\rm m}(t_{\rm e})$.
Moreover, the spectrum of runaway CR electrons does not depend on 
the magnetic field evolution except for the maximum energy of runaway CR electrons.

According to \citet{ohiraetal10,ohira11}, the energy spectrum of runaway CR electrons, 
${\rm d}N_{\rm esc}/{\rm d}E$, can be expressed by
\begin{equation}
\frac{{\rm d}N_{\rm esc}}{{\rm d}E} = A E^{-\left(s+\frac{\beta}{\alpha}\right)}~~,
\label{eq:nesc}
\end{equation}
where $A$ is the normalization factor, $s\approx2$ is the index of the energy 
spectrum of CR electrons inside the SNR and $\beta$ is a parameter to describe 
the evolution of the number of CR electrons inside the SNR 
(${\rm d}N/{\rm d}E\propto t^{\beta}E^{-s}$ where ${\rm d}N/{\rm d}E$ is the 
CR spectrum inside the SNR).
Lower-energy CRs can be produced for a longer time than high-energy CRs, 
so that the time-integrated spectrum becomes softer than the instantaneous one. 
Note that $\beta$ is not understood well for CR electrons and protons 
because their injection processes for particle acceleration have not been understood well.
In this paper, we assume that CR electrons and protons have 
the same value of $\beta=0.6$ during the Sedov phase and 
$\beta=3$ during the free expansion phase where we consider 
the thermal leakage model as the injection model \citep{ohiraetal10}. 
For $s=2.0, \alpha=2.6$ and $\beta=0.6$, the spectral index of runaway CRs 
is $s+\beta/\alpha=2.23$, which is consistent with the index expected as 
the source of Galactic CRs \citep{ohiraetal10}.
\section[]{Spatial distribution of runaway CR electrons around an SNR}
\label{sec:3}
\begin{figure}
\begin{center}
\includegraphics[width=80mm]{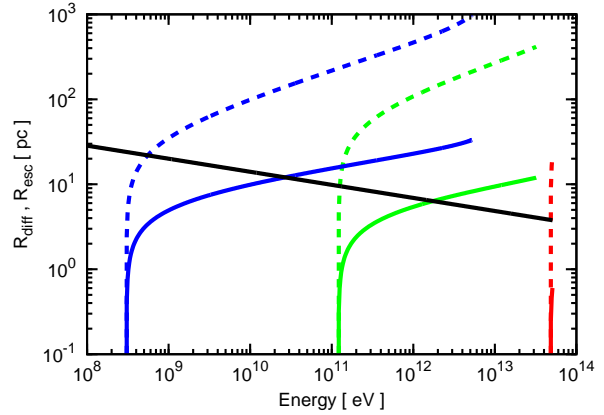}
\end{center}
\caption{Diffusion length $R_{\rm diff}$ (Equation~(\ref{eq:rd})) and 
escape radius $R_{\rm esc}$ (Equation~(\ref{eq:resc}) with $\alpha=2.6$) 
as functions of electron energy. 
The black line shows the escape radius.
The red, green, and blue lines show the diffusion length 
at $t=10^3, 10^4,~{\rm and}~10^5~{\rm yr}$, respectively.
The solid line shows the case of $\chi=0.01$ and $\delta=0.3$ and 
the dashed line shows the case of $\chi=1$ and $\delta=0.6$ where 
$\chi$ and $\delta$ are the normalization factor and 
the energy dependence of the diffusion coefficient of the ISM, respectively  (Equation~(\ref{eq:dism})).}
\label{fig:3}
\end{figure}

In this section, we derive the distribution of runaway CR electrons
at a given distance $r$ from the SNR center and at a given time $t$ from
the explosion time, $f(t,r,E)~[{\rm particles~cm^{-3}~eV^{-1}}]$.
CRs propagate around SNRs after escaping from the SNRs, 
so that CR halos are produced around the SNRs.
Suppose that the system is spherically symmetric and the diffusion
coefficient in the ISM, $D_{\rm ISM}(E)$, is spatially uniform.
Then, we solve the diffusion equation given by
\begin{equation}
\frac{\partial}{\partial t}f - D_{\rm ISM}(E)\Delta f +\frac{\partial}{\partial E}(P(E)f)= q_{\rm s}(t,E,{\bf r})~~,
\label{eq:diffusion}
\end{equation}
where $P(E)$ and $q_{\rm s}(t,E,{\bf r})$ are the energy loss rate of CR 
electrons and the source term of CR electrons, respectively.
Considering the escape process \citep{ptuskin05,ohiraetal10}, 
CRs with an energy of $E$ escape from an SNR at $t=T_{\rm esc}(E)$.  
Let ${\rm d}N_{\rm esc}/{\rm d}E$ be the total spectrum of runaway CR electrons, that is,
\begin{equation}
\frac{{\rm d}N_{\rm esc}}{{\rm d}E} = \int {\rm d}t\int{\rm d}^3r~q_{\rm s}(t,r,E) = AE^{-(s+\frac{\beta}{\alpha})}
~~.
\end{equation}
Then, the Green function, which is the solution in the case of an 
instantaneous point source,
$q_{\rm s}(t,E,{\bf r})=\delta({\bf r})\delta(t-t_{\rm esc}(E)){\rm d}N_{\rm esc}/{\rm d}E$, 
is \citep{atoyan95}
\begin{equation}
f_{\rm point}(t,r,E)=\frac{e^{-\left(\frac{r}{R_{\rm diff}(t,E)}\right)^2}}{\pi^{3/2}R_{\rm diff}(t,E)^3}\frac{P(E_0)}{P(E)}\frac{{\rm d}N_{\rm esc}(E_0)}{{\rm d}E}~~,
\label{eq:fpoint}
\end{equation}
where $E_0(t,E)$ is the initial energy defined by
\begin{equation}
\int_{E}^{E_0(t,E)} \frac{{\rm d}\epsilon}{|P(\epsilon)|}=t-T_{\rm esc}(E_0)~~,
\label{eq:e0}
\end{equation}
and the diffusion length 
while CR electrons cool from $E_0$ to $E$ is
\begin{equation}
R_{\rm diff}(E) 
= 2\left( \int_{E}^{E_0(t,E)} \frac{D_{\rm ISM}(\epsilon)}{|P(\epsilon)|}
  {\rm d}\epsilon\right)^{\frac{1}{2}}~~.
\label{eq:rd}
\end{equation}

The diffusion coefficient around an SNR has not been understood well so far. 
Although there is no guarantee that the diffusion coefficient has a power law 
form, $D_{\rm ISM}\propto E^{\delta}$, \citep[e.g.][]{fujita10,fujita11}, 
such a diffusion coefficient can explain gamma-ray spectra of 
middle-aged SNRs observed by {\it Fermi} \citep[e.g.][]{ohiraetal11,li11}.
Therefore, we assume the diffusion coefficient to be
\begin{equation}
D_{\rm ISM}(E)=10^{28}~\chi \left( \frac{E}{10~{\rm GeV}}\right)^{\delta} {\rm cm^2~s^{-1}}~~.
\label{eq:dism}
\end{equation}
The values of $\delta$ and $\chi$ are uncertain. 
The Galactic mean values are 
$\delta\approx0.3-0.6$ and $\chi\approx1$ \citep{berezinskii90}.
However,  $\chi$ is expected to be much smaller than unity ($\chi\approx0.01$) 
around SNRs 
because CRs amplify magnetic fluctuations around SNRs
\citep{fujita09,fujita10,fujita11,giuliani10,torres10,li10}.

We find that the effect of finite source size is important. 
To see this,
we show, in Figure~\ref{fig:3}, the diffusion length $R_{\rm diff}(E)$
(color lines) and the escape radius $R_{\rm esc}(E)$ (black solid line);
the former is obtained from Equations~(\ref{eq:tesc}), (\ref{eq:e0}) and
(\ref{eq:rd}), while the latter is
obtained from Equation (\ref{eq:resc}),
where we consider synchrotron cooling with $B=3~{\rm \mu G}$ and inverse
Compton cooling with the Galactic radiation field provided by the
$8~{\rm kpc}$ model of \citet{porter08}.
We fully consider the Klein-Nishina effect for inverse Compton scattering \citep{blumenthal70}.
As shown in Figure~\ref{fig:3}, $R_{\rm diff}(E)$
has low- and high-energy cutoffs.
The lower cutoff corresponds to the energy of just escaping 
CR electrons at the SNR age, $t=t_{\rm esc}(E_0)$.
The higher cutoff shows the maximum energy of runaway 
CR electrons at the SNR age $t$. 
It is found that
for $t=10^3~{\rm yr}$ and $\chi=0.01$ (red solid line), CR electrons
above $50~{\rm TeV}$ escape from the SNR and the diffusion length of
runaway CR electrons is smaller than the escape radius.
Similarly, 
for $t=10^4~{\rm yr}$ and $\chi=0.01$ (green solid line), CR electrons
above $\sim 100~{\rm GeV}$ escape from the SNR and the diffusion
length of runaway CR electrons with a few TeV is comparable to the
escape radius.
Therefore, the point source approximation is not well, that is, 
the finiteness of the source size is important.
We also remind that even if SNRs produce CR electrons 
with $100~{\rm TeV}$, the SNRs should be younger than 
about $10^4~{\rm yr}$ (cooling time) and 
locate within about $500~{\rm pc}$ from the Earth to observe the 
CR electrons directly.

Now, taking into account the fact that CRs escape from the SNR surface \citep{ohiraetal10}, the source term is replaced with
\begin{equation}
q_{\rm s}=\frac{1}{4\pi r^2}\frac{{\rm d}N_{\rm esc}}{{\rm d}E} \delta(r-R_{\rm esc}(E)) \delta(t-T_{\rm esc}(E))~~,
\end{equation}
where $R_{\rm esc}(E)$ is the radius where the CRs 
with an energy of $E$ escape from the SNR.
Then, we find the solution to equation~(\ref{eq:diffusion}) as \citep{ohiraetal11}
\begin{eqnarray}
f_{\rm ext}(t,r,E)&=&\frac{e^{-\left(\frac{r-R_{\rm esc}(E_0)}{R_{\rm diff}(t,E)}\right)^2}-e^{-\left(\frac{r+R_{\rm esc}(E_0)}{R_{\rm diff}(t,E)}\right)^2}}{4\pi^{3/2}R_{\rm diff}(t,E)R_{\rm esc}(E_0)r} \nonumber \\
&& \nonumber \\
&&\times\frac{P(E_0)}{P(E)}\frac{{\rm d}N_{\rm esc}(E_0)}{{\rm d}E}~~.
\label{eq:fext}
\end{eqnarray}
The finiteness of the source size is important 
in the vicinity of the escape radius ($r\approx R_{\rm esc}$)
for $R_{\rm esc}\gg R_{\rm diff}$.
This means $r-R_{\rm esc} \ll R_{\rm diff}$ and $r+R_{\rm esc} \gg R_{\rm diff}$, 
so that $f_{\rm ext}$ is approximately given by \citep{ohiraetal11}
\begin{equation}
f_{\rm ext}(t,r,E)\approx \frac{1}{4\pi^{3/2}R_{\rm diff}(t,E)R_{\rm esc}(E_0)r} \frac{P(E_0)}{P(E)}\frac{{\rm d}N_{\rm esc}(E_0)}{{\rm d}E}~~.
\end{equation}
%

\section[]{Radiation spectra due to runaway CR electrons from SNRs}
\label{sec:4}
\begin{figure}
\begin{center}
\includegraphics[width=80mm]{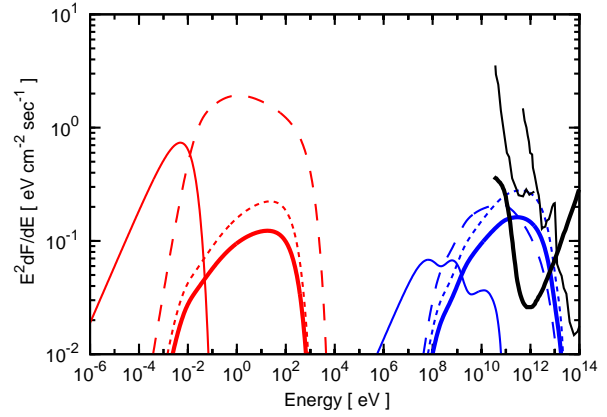}
\end{center}
\caption{Radiation spectra for $d=3~{\rm kpc}, U_{\rm CR,e}=10^{48}~{\rm erg}, 
t=10^4~{\rm yr}, \chi=0.01$, and $\delta=0.6$ where $\chi$ and $\delta$ are 
the normalization factor and the energy dependence of the diffusion coefficient 
of the ISM, respectively. 
The red and blue lines show synchrotron radiation and 
inverse Compton scattering, respectively. 
The thick solid line shows a contribution from runaway CR electrons 
outside the SNR ($R_{\rm sh}<\rho \leq \rho_{\rm size}$) 
(Equation~(\ref{eq:nprojout})), the thin solid line shows a contribution from 
trapped CR electrons which have not escaped (Equation~(\ref{eq:ntrap})), 
the long dashed line shows a contribution from runaway CR electrons which 
have been caught up by the shock (Equation~(\ref{eq:ncaught})), 
and short dashed lines show a contribution from runaway CR electrons which  
have not been caught up by the shock but are inside the SNR projected on the 
sky ($\rho \leq R_{\rm sh}$)  (Equation~(\ref{eq:nprojin})), respectively. 
The thick and thin black lines show CTA $50~{\rm h}$ \citep{cta10} and 
LHAASO (where the MAGIC-II-type telescopes improves the low energy 
LHAASO sensitivity) \citep{cao10} integral sensitivities of point sources, respectively.
}
\label{fig:4}
\end{figure}
\begin{figure}
\begin{center}
\includegraphics[width=80mm]{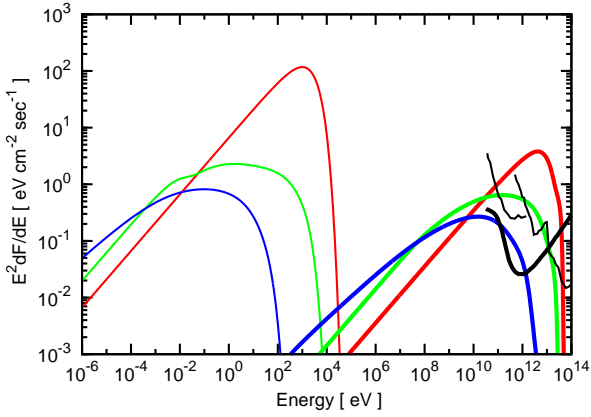}
\end{center}
\caption{Time evolution of radiation spectra from the SNR 
($0\leq \rho \leq \rho_{\rm size}=1.2 R_{\rm sh}$) 
for $d=3~{\rm kpc}, U_{\rm CR,e}=10^{48}~{\rm erg}, 
\chi=0.01$ and $\delta=0.6$. 
The thin and thick color lines show total synchrotron spectra and 
total inverse Compton spectra, respectively.
The red, green, and blue lines correspond to spectra at the age of $10^3, 10^4$, 
and $10^5~{\rm yr}$, respectively.
The thick and thin black lines show CTA $50~{\rm h}$ \citep{cta10} and LHAASO (where the MAGIC-II-type telescopes improves the low energy LHAASO sensitivity) \citep{cao10} integral sensitivities of point sources, 
respectively.}
\label{fig:5}
\end{figure}
\begin{figure}
\begin{center}
\includegraphics[width=80mm]{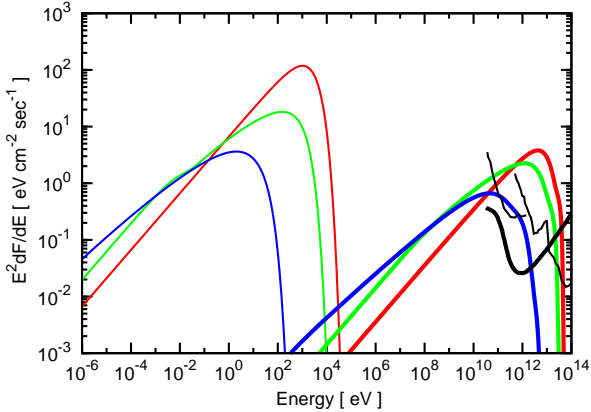}
\end{center}
\caption{The same as Figure~\ref{fig:5}, but for $\delta=0.3$.}
\label{fig:6}
\end{figure}
\begin{figure}
\begin{center}
\includegraphics[width=80mm]{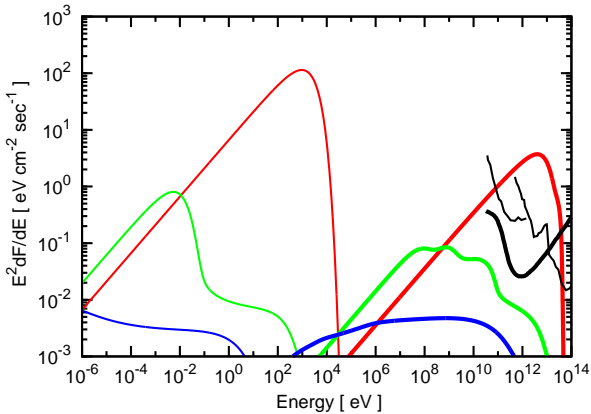}
\end{center}
\caption{The same as Figure~\ref{fig:5}, but for $\chi=1$.}
\label{fig:7}
\end{figure}
In this section, radiation spectra from runaway CR 
electrons are calculated. 
They should be calculated using their volume-integrated distribution function. 
At first, we consider the energy spectrum of runaway CR electrons
contributing to the radiation spectrum, ${\rm d} N_{\rm proj}/{\rm d} E$.
Emission far outside the SNR is too dim to be detected 
because the number of emitting electrons is too small.
Taking into account an observation sensitivity,
we should consider the projected distance $\rho_{\rm size}$ from 
the SNR center on the sky.
We adopt $\rho_{\rm size} = 1.2R_{\rm sh}$.
Then, we obtain
\begin{equation}
\frac{{\rm d} N_{\rm proj}}{{\rm d} E}=\int_{0}^{\rho_{\rm size}} {\rm d}\rho ~2\pi \rho \int_{-\infty}^{\infty}{\rm d}z ~f_{\rm ext}(t,\sqrt{z^2+\rho^2},E) ~~,
\label{eq:f}
\end{equation}
where the $z$-direction is along the line of sight.
In the energy region satisfying the condition $\rho \ll R_{\rm diff}(t,E)$, 
the $z$-integration becomes approximately 
$f_{\rm point}R_{\rm diff}$.
Then, ${\rm d} N_{\rm proj}/{\rm d} E$ can be written by
\begin{eqnarray}
\frac{{\rm d} N_{\rm proj}}{{\rm d} E}&\propto& \left(\frac{\rho_{\rm size}}{R_{\rm diff}(t,E)}\right)^2\frac{{\rm d}N_{\rm esc}}{{\rm d}E} \nonumber \\
&\propto& \chi^{-1}E^{-\left(\delta+s+\frac{\beta}{\alpha}\right)} ~~,
\label{eq:fapp}
\end{eqnarray}
where radiative losses are neglected. 

We further divide runaway CR electrons into three groups as in the following: 
\begin{itemize}
\item
The energy spectrum of runaway CR electrons outside the SNR projected on the sky ($R_{\rm sh}<\rho \leq \rho_{\rm size}$), ${\rm d} N_{\rm proj,out}/{\rm d} E$,  is obtained by
\begin{equation}
\frac{{\rm d} N_{\rm proj,out}}{{\rm d} E}=\int_{R_{\rm sh}}^{\rho_{\rm size}} {\rm d}\rho ~2\pi \rho \int_{-\infty}^{\infty}{\rm d}z ~f_{\rm ext}(t,\sqrt{z^2+\rho^2},E) ~~,
\label{eq:nprojout}
\end{equation}

\item
Some of runaway CR electrons are caught up by the SNR shock 
because the SNR shock expands in the halo of runaway CR electrons.
Note that most runaway CR electrons are outside the SNR 
because the diffusion length increases with time faster than the SNR shock.
The energy spectrum of runaway CR electrons which have been caught 
up by the shock, ${\rm d} N_{\rm caught}/{\rm d} E$, is obtained by
\begin{equation}
\frac{{\rm d} N_{\rm caught}}{{\rm d} E}=\int_0^{R_{\rm sh}} {\rm d}r~4\pi r^2 f_{\rm ext}(t,r,E)~~.
\label{eq:ncaught}
\end{equation}

\item
The energy spectrum of runaway CR electrons which have not been caught up by the shock 
but are inside the SNR projected on the sky ($\rho \leq R_{\rm sh}$), ${\rm d} N_{\rm proj,in}/{\rm d} E$,  is obtained by
\begin{equation}
\frac{{\rm d} N_{\rm proj,in}}{{\rm d} E}= \frac{{\rm d} N_{\rm proj}}{{\rm d} E} -\frac{{\rm d} N_{\rm proj,out}}{{\rm d} E}- \frac{{\rm d} N_{\rm caught}}{{\rm d} E}~~,
\label{eq:nprojin}
\end{equation}

\end{itemize}

Finally, we consider trapped CR electrons.
CR electrons with an energy smaller than 
the escape-limited maximum energy $E_{\rm m,esc}(t)$ 
are still trapped in the SNR.
Their energy spectrum for $t>t_{\rm e}$, ${\rm d}N_{\rm trap}/{\rm d}E$ is given by
\begin{equation}
\frac{{\rm d}N_{\rm trap}}{{\rm d}E}=A\left( E_{\rm m,esc}\left(t\right) \right)^{-\left(s+\frac{\beta}{\alpha}\right)}\left(\frac{E}{E_{\rm m,esc}(t)}\right)^{-s}~~.
\label{eq:ntrap}
\end{equation}

In order to calculate each energy spectrum described above,
several model parameters in section~2 and 3 are fixed.
Throughout this section, we only consider the model of 
$\eta_{\rm g}=\eta_{\rm g,free}$ for the 
magnetic field evolution, that is, {the start time of escape of CR electrons}  
is $t_{\rm e}=9.70\times10^2~{\rm yr}$.
In our formulation shown in Section~{\ref{sec:2}},
the spectral index of runaway 
CR electrons does not depend on models of the magnetic field 
evolution, while  the maximum energy of runaway CR electrons, $E_{\rm m,e}$, 
and the start time of escape of CR electrons, $t_{\rm e}$, do depend.
We adopt $R_{\rm S}=2~{\rm pc}$, 
$t_{\rm S}=200~{\rm yr}$, $B_{\rm ISM}=3~{\rm \mu G}$,
$\eta_{\rm acc}=10$, $s=2.0$, $\alpha=2.6$, $\beta=0.6$, 
$d=3~{\rm kpc}$, $n_{\rm ISM}=1~{\rm cm}^{-3}$, 
and $U_{\rm CR,e}=10^{48}~{\rm erg}$, where $d$, $n_{\rm ISM}$, 
and $U_{\rm CR,e}$ are the source distance, 
the number density of the ISM, 
and the total energy of runaway CR electrons 
from $1~{\rm GeV}$ to $E_{\rm knee}$, respectively.
We calculate inverse Compton scattering 
with the Galactic radiation 
field provided by the $8~{\rm kpc}$ model of \citet{porter08}
and synchrotron radiation.
We fully consider the Klein-Nishina effect for inverse Compton 
scattering \citep{blumenthal70}. 
We use the downstream magnetic field $B_{\rm d}=4B$ for 
synchrotron radiation emitted inside the SNR.

For given distributions of CR electrons 
(Equations~(\ref{eq:nprojout})--(\ref{eq:ntrap})), radiation spectra are calculated.
Figure~\ref{fig:4} shows the spectral components 
of four CR electron groups given by Equations~(\ref{eq:nprojout})--(\ref{eq:ntrap}),
where  $t=10^4~{\rm yr}~(>t_{\rm e})$, $\chi=0.01$, and $\delta=0.6$
($\chi$ and $\delta$ are the normalization factor and the energy dependence of 
the diffusion coefficient: see Equation~(\ref{eq:dism})).
We also show in the figure the integral sensitivities of point sources of 
the future TeV gamma-ray telescopes, 
CTA $50~{\rm h}$ \citep{cta10} and LHAASO \citep{cao10} (black lines).
The thin and thick lines show the radiation spectra coming  from the interior 
of the SNR region projected onto the sky (that is $\rho \leq R_{\rm sh}$) 
and from the exterior of the SNR 
($R_{\rm sh}<\rho \leq \rho_{\rm size}=1.2R_{\rm sh}$) (Equation~(\ref{eq:nprojout})), respectively. 
The thin-solid lines show spectra emitted by trapped CRs 
which have not escaped from the SNR yet (Equation~(\ref{eq:ntrap})). 
The long and short-dashed lines show the spectra 
emitted by runaway CR electrons which have been caught up by the 
shock (Equation~(\ref{eq:ncaught})) and have not been caught up 
by the shock (Equation~(\ref{eq:nprojin})), respectively.

One can see from Figure~\ref{fig:4} that 
maximum-energy photons of synchrotron radiation are emitted by 
runaway CR electrons which have been caught up by the shock. 
This is completely different from the standard picture that maximum 
energy synchrotron photons are produced by CR electrons accelerating 
at the SNR shock.
Note that this new picture can be applied after CR electrons start to 
escape ($t>t_{\rm e}$). 
In previous works \citep[e.g.][]{berezhko10,zirakashvili10,edmon11}, 
synchrotron emission comes from CR electrons advected 
downstream at any time, that is, there is no halo of runaway CRs.
In our model, the diffusion coefficient rapidly increases with time 
compared with the previous work.
Therefore, in our model, CR electrons can escape from inside SNRs 
over advection and the halo of runaway CRs is produced.

Radiation spectra from runaway CRs are softer than that 
from trapped CRs. 
This is because higher-energy CRs are more diluted by diffusion 
and the spectrum of runaway CRs becomes softer than that of 
trapped CRs (see Equation~(\ref{eq:nesc})).
From Equations~(\ref{eq:fpoint}) and (\ref{eq:ncaught}), 
the energy spectrum of caught-up CR electrons can be written by
\begin{eqnarray}
\frac{{\rm d} N_{\rm caught}}{{\rm d} E}&\propto& \left( \frac{R_{\rm sh}}{R_{\rm diff}}\right)^3 \frac{{\rm d}N_{\rm esc}}{{\rm d}E} \nonumber \\
&\propto&\chi^{-\frac{3}{2}}E^{-\left(s+\frac{\beta}{\alpha}+1.5\delta \right)}  \nonumber \\
&\propto&E^{-3.13}~~.
\label{eq:ncaught2}
\end{eqnarray}
where we assume $s=2.0, \beta=0.6, \alpha=2.6$ and $\delta=2.6$ 
at the last equation.
Therefore, the energy spectrum of synchrotron radiation from 
caught-up CR electrons becomes flat (long-dashed line). 
In addition, we predict that inverse Compton scattering outside the middle-aged 
SNR (thick blue line) will be observed by CTA and LHAASO if the SNR produces 
sufficient CR electrons ($U_{\rm CR,e}=10^{48} ~{\rm erg}$) and $\chi=0.01$.

In Figures~\ref{fig:5}--\ref{fig:7}, we show time 
evolutions of radiation spectra around the SNR 
($\rho<\rho_{\rm size}=1.2R_{\rm sh}$) for three parameter sets of 
the diffusion coefficient of the ISM, 
$(\chi,\delta)=(0.01,0.6), (0.01,0.3)$, and $(1,0.6)$.
Integral sensitivities of point sources of the future ground-based 
Cherenkov telescopes, 
CTA $50~{\rm h}$ \citep{cta10} and LHAASO \citep{cao10} (black lines), 
are also shown.
The thick solid and thin solid lines show total synchrotron radiation and
 total inverse Compton scattering, respectively.
Figures~\ref{fig:5} and \ref{fig:6} (small diffusion coefficient case) show 
that even though the SNR age is $10^5~{\rm yr}$, the future 
Cherenkov telescopes can detect radiation from runaway CR 
electrons.

Figures~\ref{fig:5} and \ref{fig:7} show that radiation from runaway 
CR electrons depends on the normalization of the diffusion coefficient, $\chi$.
As shown in Equation~(\ref{eq:fapp}), the flux of inverse Compton scattering 
is large for small $\chi$.
Therefore, if $\chi<0.1$, radiation due to runaway CRs from
middle-aged and old SNRs can be potentially observed by CTA and LHAASO.
Similarly, as shown in Equation~(\ref{eq:ncaught2}), the flux of synchrotron 
radiation from caught-up CR electrons is large for small $\chi$. 
It should be noted that if the evolution of the magnetic field is 
$B^2 \propto u_{\rm sh}^3$ or $u_{\rm sh}^2$,
the maximum energy of caught-up CR electrons is too small to 
produce synchrotron X-ray photons (Equation~(\ref{eq:eescmax})).
However, if the magnetic field at the SNR shock is suddenly amplified 
by interactions of molecular clouds \citep{inoue09},
X-ray photons could be produced by synchrotron radiation 
from caught-up CR electrons. 
\section{Discussion}
\label{sec:5}
\begin{figure}
\begin{center}
\includegraphics[width=80mm]{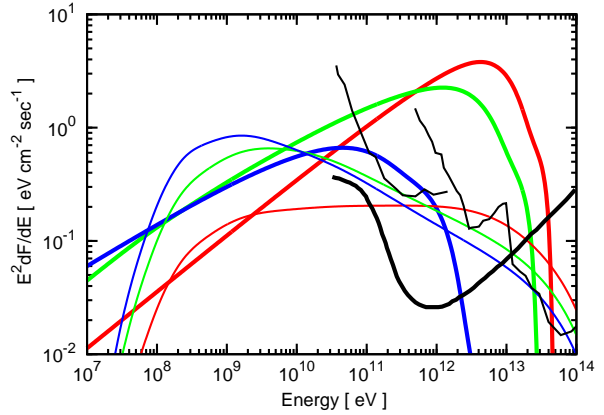}
\end{center}
\caption{Time evolution of gamma-ray spectra from the SNR 
(with the projected distance $\rho_{\rm size}<1.2 R_{\rm sh}$) 
for $U_{\rm CR,p}=10^{50}~{\rm erg}, n_{\rm ISM}=1~{\rm cm}^{-3}, \chi=0.01, \delta=0.3$, and other parameters are the same as Figure~\ref{fig:5}.
The thick and thin lines show the total inverse Compton spectra and the gamma rays ($\pi^0\rightarrow2\gamma$) from runaway CR protons, respectively.
The red, green, and blue lines correspond to spectra at the age of $10^3, 10^4$, 
and $10^5~{\rm yr}$, respectively.
The thick and thin black lines show CTA $50~{\rm h}$ \citep{cta10} and LHAASO \citep{cao10} sensitivities, respectively.}
\label{fig:8}
\end{figure}

In this paper, taking into account escape of CR electrons from SNRs,
we have investigated the evolution of 
the maximum energy of CR electrons, as well as that of
the magnetic field and the maximum energy of CR protons.
We found the followings:
\begin{itemize}
\item CR electrons start to escape from SNRs later than CR nuclei (Equation~(\ref{eq:te})). 
\item SNRs can produce CR electrons up to about $0.3-50~{\rm TeV}$, 
which depends on the evolution of the magnetic field 
(Equation~(\ref{eq:eescmax})).
\item If the magnetic field decays slowly, 
the maximum energy of runaway CR electrons becomes small (Equation~(\ref{eq:eescmax})).
\item The maximum energy of CR electrons is anticorrelated with that of CR nuclei 
(Equation~(\ref{eq:eescmax})).
\item The spectrum of runaway CR electrons is unaffected by the cooling break 
of synchrotron radiation (Section~(\ref{sec:2.6})).
\end{itemize} 
In addition, we calculated the spatial distribution of runaway 
CR electrons around SNRs.
Using this distribution, we calculated synchrotron radiation and 
inverse Compton scattering from the runaway CR electrons 
around SNRs. 
We found the followings:
\begin{itemize}
\item Inverse Compton radiation of the runaway CR 
electrons can be observed by future ground-based Cherenkov telescopes, 
such as CTA and LHAASO.
Because the cosmic microwave background is inevitably present, CTA 
and LHAASO can impose tight constraint on the amount of CR electrons.
\item Maximum-energy photons of synchrotron radiation are emitted by 
runaway CR electrons which have been caught up by the shock.
\item Radiation spectra from runaway CR electrons depend on the diffusion coefficient of ISM.
\end{itemize}

Although there are many parameters in determining the evolution
of the maximum energy
($R_{\rm S}$, $t_{\rm S}$, $B_{\rm ISM}$, $E_{\rm knee}$,
$\eta_{\rm acc}$, $\eta_{\rm esc}$, $\eta_{\rm g,free}$,
$\alpha$ and $\alpha_{\rm B}$; see Table~1),
all except $\alpha_{\rm B}$ have typical or appropriate values.
The value of $\alpha_{\rm B}$, which is a parameter to describe 
the evolution of the magnetic field, is highly uncertain.
We have derived some relations for the general $\alpha_{\rm B}$ and 
especially discussed three cases.
We have assumed that the end of the Sedov phase is  
$200~{\rm yr} \times 10^{2.5} \sim 63~{\rm kyr}$.
On the other hand, \citet{truelove99} and \citet{petruk05} thought that it is 
$12~{\rm kyr}$ and $30~{\rm kyr}$, respectively. 
Even when we take these values, $12$ or $30~{\rm kyr}$, 
our results do not change significantly.
In this paper, we have considered only one fiducial model of SNRs 
($t_{S}=200~{\rm yr}$ and $R_{\rm S}=2~{\rm pc}$).
It is also an interesting future work to examine the individual SNR
detected by H.E.S.S., MAGIC, and VERITAS.
It should be noted that the age of most young SNRs is about 
the beginning of the Sedov phase and smaller than {the start time of escape of CR electrons}, that is, $t\sim t_{\rm S}<t_{\rm e}$ 
\citep[see the table 1 of][]{truelove99}.
Therefore, CR electrons have not escaped from those SNRs.

We consider three evolution models of the magnetic field in 
Section~\ref{sec:2}.
Different models predict different evolutions of the maximum 
energy limited by synchrotron cooling during 
$t_{\rm S}<t< \min \{t_{\rm e},t_{\rm B}\}$. 
Maximum photon energies of synchrotron radiation and 
inverse Compton scattering depend on evolution models of 
the magnetic field.
Furthermore, the maximum energy of runaway CR electrons also 
depends on the evolution model of the magnetic field ($0.3-50~{\rm TeV}$).
Figures~\ref{fig:5}, \ref{fig:6}, and \ref{fig:7} show that CTA and 
LHAASO can observe the maximum value. 
Hence, CTA and LHAASO can provide constraints on evolution models of the 
magnetic field.
To make a more accurate radiation spectrum, we should 
consider escaping and trapped CR spectra in detail \citep{caprioli09,reville09}
and solve SNR dynamics and the diffusion-convection equation \citep{zirakashvili10}.

Fermi and H.E.S.S. show the spectral break or the cut off of the CR 
electron spectrum at around $1~{\rm TeV}$.
The magnetic field evolution model of $B^2\propto u_{\rm sh}^3$ 
could explain the break as the maximum energy of CR electrons that 
can escape from SNRs (see Equation~(\ref{eq:eescmax})).
This model is suggested by other observations \citep{vink08} and 
by \citet{bell04}.
A local source may contribute to high energy CR electrons and 
the maximum energy of CR electrons may be limited by cooling during 
the propagation \citep[e.g.][]{kobayashi04,thoudam12}. 
Even so, we can rule out any models that predict a smaller maximum energy 
of CR electrons than that of observed CR electrons. 
A more tight constraint on the magnetic field evolution will be given by future 
observations of the CR electron spectrum, AMS-02 and CALET.
To make a more realistic prediction, we should perform 
time-dependent fluid simulations \citep{schure10} 
and consider the CR back reaction \citep[e.g.][]{drury89,ellison90,berezhko97,blasi02,kang02}.

In this paper, we have focused on escape of accelerated electrons. 
SNRs produce CR protons as well as CR electrons, so that
the hadronic gamma-ray radiation due to the neutral pion decay should be
compared to the leptonic gamma rays.
In figure~\ref{fig:8}, we show an example of the gamma-ray spectrum.
We adopt $U_{\rm CR,p}=10^{50}~{\rm erg}, n_{\rm ISM}=1~{\rm cm}^{-3}, 
\chi=0.01, \delta=0.3$, and otherparameters are the same as in the previous section.
We use the code provided by \citet{kamae06,karlsson08}
\footnote[1]{
Note that in Figure~\ref{fig:8}, the softening around $30~{\rm TeV}$ 
of gamma rays from CR protons is due to the limitation of the code 
provided by \citet{kamae06,karlsson08}.
The code calculates gamma-ray spectra 
from CR protons with energies up to $512~{\rm TeV}$.
}
to obtain gamma-ray spectra from CR protons.
Note that if the ratio of CR electrons to CR protons is 
$U_{\rm CR,e}/U_{\rm CR,p}=10^{-2}$ and the number density of 
the ISM is $n_{\rm ISM}=1~{\rm cm}^{-3}$, 
inverse Compton scattering due to runaway CR electrons dominates over 
the hadronic component at the TeV energy region.
The maximum photon energy and its flux decrease with time for 
synchrotron radiation and inverse Compton scattering because of 
cooling and diffusion.
On the other hand, the maximum energy of gamma rays from CR 
protons does not change because the cooling time of CR protons 
is longer than the SNR age.
Therefore, radiation from CR protons can dominate over that from CR 
electrons at the $100~{\rm TeV}$ energy region for middle-aged and/or old SNRs.

We comment on a possible origin of unidentified 
very-high-energy gamma-ray
sources (so called, TeV-unIDs) discovered by H.E.S.S.
\citep{aharonian05,aharonian06b,aharonian08}.
Figures~\ref{fig:5}, \ref{fig:6}, and \ref{fig:7} tell us that 
old SNRs ($t\sim10^{5}~{\rm yr}$) could be the origin of TeV-unIDs.
These figures show that inverse 
Compton scattering from runaway CR electrons can be observed 
in the TeV region, but synchrotron radiation can not produce X-ray 
photons for old SNRs.
The maximum energy of runaway CR electrons decreases with time 
owing to synchrotron and inverse Compton cooling and the 
maximum energy is a few TeV at $t=10^5~{\rm yr}$
\citep[see also][]{yamazaki06,ioka2010}.
Electrons with energies of a few TeV can produce TeV gamma-rays 
by inverse Compton scattering with optical photons, but can not 
produce X-ray photons by synchrotron radiation with the magnetic 
field of $3~{\rm \mu G}$. 
This is the same reason why pulsar wind nebulae (PWNe) are plausible 
candidates for TeV-unIDs \citep{dejager08}.
Radiation from runaway CR electrons depends on the number of CR electrons, 
the diffusion coefficient around the SNR, the SNR age, and the source distance.
For example, if the value of $\chi$ is larger than 0.1, 
the flux of inverse Compton scattering from runaway CR electrons around an SNR
older than $10^{4}~{\rm yrs}$ with 
the distance $3~{\rm kpc}$ is smaller than the sensitivity of H.E.S.S.,
so that the number of currently detected TeV unIDs is not so large.
However, even in this case, we expect that future observations by CTA 
will increase the number.

In addition, we comment on escape of CR electrons from PWNe.
\citet{kawanaka11} investigated CR electron spectra from a young pulsar 
embedded in the SNR, but did not take into account synchrotron cooling 
while CR electrons cross the SNR shell. 
If the diffusion escape time from the SNR is longer than the cooling time,
\citet{kawanaka11} is not valid.
However, CR electrons are not produced at SNR shocks for PWNe.
That is, the maximum energy of CR electrons produced in PWNe could be 
larger than that produced at SNR shocks.
In this case, CR electrons produced in PWNe can escape from 
SNRs as long as the diffusion escape time is smaller than the cooling time.
This depends on the evolution of the magnetic field.
\citet{kawanaka11} proposed that 
the low energy cutoff of CR electrons escaping from nearby sources 
could be observed in the above TeV band.
Therefore, \citet{kawanaka11} is complementary to this paper that 
investigated the high energy cutoff.

Finally, we emphasize that 
escape of accelerated particles is common in other high energy particle accelerators 
(e.g. microquasars, gamma ray bursts and active galactic nuclei), 
so that radiation of runaway accelerated particles from the accelerators can also be expected.

\section*{Acknowledgments}

The authors thank A. Bamba, E. A. Helder and K. M. Schure for useful 
discussions and comments. 
We also thank the referee for valuable comments to improve the paper.
This work is supported in part by grant-in-aid from the Ministry of
Education, Culture, Sports, Science, and Technology (MEXT) of Japan, 
No.~24$\cdot$8344(Y.~O.), No.~21740184 (R.~Y.), Nos.~22244019, 
22244030 (K.~I.), No.~22740131 (N.~K.), No.~21684014 (Y.~O. and K.~I.), 
No.~19047004 (R.~Y. and K.~I.), and ERC advanced research grant (N.~K.).

\bsp
\label{lastpage}
\end{document}